\documentclass[showpacs,preprint,aps]{revtex4}
\usepackage{graphicx}% Include figure files
\usepackage{dcolumn}% Align table columns on decimal point
\usepackage{bm}% bold math
\begin{document}
\setcounter{page}{1}
\title
{Tunneling times and bremsstrahlung in alpha decay}
\author
{N. G. Kelkar and M. Nowakowski}
\affiliation{ Departamento de Fisica, Universidad de los Andes,
Cra.1E No.18A-10, Santafe de Bogota, Colombia}
\begin{abstract}
A semi-classical model based on quantum time concepts is 
presented for the evaluation of bremsstrahlung emission probabilities in 
alpha decay of nuclei. 
The contribution to the bremsstrahlung emission from the different regions 
in tunneling is investigated using realistic double folded nuclear and 
Coulomb potentials. Within this model, the contribution from the 
radiation emitted in front of the barrier before tunneling is much larger than 
that while leaving the barrier. 
A comparison with the data on $^{210}$Po shows that 
the results are sensitive to the nuclear potential 
and the rectangular well used in many of the quantum mechanical approaches can 
even give qualitatively different results. 
\end{abstract}
% these are paxnumbers of the alpha decay dwell time paper
\maketitle

\section{Introduction}
The emission of photons accompanying the Coulomb interaction of 
charged particles is well explained by classical electrodynamics. 
The strength of the electromagnetic radiation is proportional to the 
acceleration which the charged particle experiences in an external field. 
In order to study bremsstrahlung emission accompanying alpha decay in 
nuclei, however, one needs to go beyond the classical picture where an 
alpha particle is accelerated in the Coulomb field of the daughter nucleus. 
In contrast to the photon emission accompanying nuclear beta decay, 
the photons in alpha decay can also be emitted during the quantum tunneling 
process. The natural question that arises is therefore: do the $\alpha$ 
particles emit radiation during tunneling or do they emit only in their 
acceleration outside the barrier? 
This curiosity gave rise to experiments measuring the emission probabilities 
of photons in the alpha decay of $^{214}$Po \cite{arrigo,giard1}, 
$^{210}$Po \cite{kasagi1, boie, eremin}, 
$^{226}$Ra \cite{arrigo,giard2} and $^{244}$Cu 
\cite{kasagi2}. However, with the emission probabilities being small and 
the experiments difficult to perform, there remained discrepancies in data. 
The theoretical calculations trying to explain these data also saw a 
similar fate. For example, 
the authors in \cite{kasagi1} used an existing 
theoretical approach \cite{dyak1} based on a semi-classical calculation 
of the tunneling motion through the barrier and found very good agreement 
with their data. A repetition of the same calculation in a different manner 
\cite{dyak2}, however, generated qualitatively different results. 
In \cite{papen}, within a fully quantum mechanical approach, the authors 
found that the main contribution to photon emission arose from Coulomb 
acceleration and the under barrier tunneling contribution was tiny. 
The authors in \cite{takigawa} however concluded that the total contribution 
results from a subtle interference of the tunneling, mixed 
and classical regions. Different aspects of this process, such as a 
time dependent description \cite{berty}, the ``interference of space 
regions" \cite{taklya}, analysis of angular bremsstrahlung spectra 
\cite{madan}, the dynamic characteristics 
such as the position, velocity and acceleration of the $\alpha$ particle 
\cite{greiner}, contribution of quadrupole radiation \cite{jent}  
etc have also been studied. However, with the lack of data,
the discrepancies in the understanding of the bremsstrahlung emission 
in alpha decay remain. The present work attempts to analyze some of the 
issues with a new semi-classical approach based on tunneling times.

In the next section, after a brief introduction to the time concepts 
used in the present work, we shall present a semi-classical model to evaluate 
the photon emission probabilities in alpha decay. In particular we 
consider the case of alpha decay in $^{210}$Po. Though some of the theoretical 
approaches in literature perform a fully quantum mechanical treatment of the 
problem, 
not much attention is paid to the details of the nuclear potential.
We present results displaying the sensitivity of the calculations to 
the nuclear potential used, the necessity of including an alpha 
cluster preformation factor and the role of the under barrier and 
outside the barrier acceleration of the alpha particle.
Finally, before summarising our results, we present a section with a 
critical view of the various theoretical approaches available. 

\section{Tunneling times}
Tunneling is one of the most remarkable phenomenon of quantum physics. 
Interesting is also the question of how long does a particle take 
to traverse the barrier. The latter indeed gave rise to several quantum 
time concepts such as the phase, dwell, traversal and Larmor time 
\cite{muga}. With the availability of so many definitions (which 
some times even include complex times \cite{sokol,hauge}), it is of interest 
to inspect which of these times could correspond to physically measured 
quantities. The stationary concepts of dwell time and traversal time do find 
a connection with measurable quantities, 
with the former giving the half life of radioactive nuclei 
and the latter the inverse of the assault frequency in alpha particle 
tunneling \cite{us1}. It is these two concepts which we shall use below in 
developing a semi-classical model for bremsstrahlung in alpha decay.
Before going over to the model, we briefly introduce the two concepts.
  
Given an arbitrary potential barrier $V(x)$ in one-dimension (a framework
which is also suitable for spherically symmetric problems),
confined to an interval
$(x_1, x_2)$, the dwell time is given by the number of particles in
the region divided by the incident flux $j$:
\begin{equation}\label{dwelltime}
\tau_D\,=\,{\int_{x_1}^{x_2}\,|\Psi(x)|^2\,dx \over j}\, .
\end{equation}
Here $\Psi(x)$ is the time independent solution of the Schr\"odinger equation
in the given region.
The dwell time is usually defined as the time spent in the region
$(x_1, x_2)$ regardless of how the particle escaped (by reflection or
transmission) and $j = \hbar \,k_0 \,/\mu$ 
(where $k_0 = \sqrt{2 \mu E} / \hbar$ with $E$ being
the kinetic energy of the tunneling particle and $\mu$ the reduced mass) 
for a free particle.
In case that one defines the dwell time for a particle 
bound in a region which either got transmitted or reflected later, 
the flux $j$ gets replaced by the transmitted or
reflected fluxes,
$j_T = \hbar \, k_0 |T|^2/ \mu$ and $j_R = \hbar \, k_0 |R|^2 / \mu$
\cite{mario, us1} respectively. 
Here $|T|^2$ and $|R|^2$ are the transmission and reflection coefficients
(with $|T|^2 \, +\, |R|^2 \,=\, 1$ due to conservation of probability). 
The traversal time defined by B\"uttiker \cite{butik} is somewhat different
and is given as,
\begin{equation}\label{travtime}
\tau_{trav} (E)\,=\,\int_{x_1}^{x_2}\,\,{\mu \over \hbar \, k(x)}\,\,dx\, ,
\end{equation}
where, 
$k(x)\,=\,{\sqrt{ 2\mu\,(|V(x)\,-\,E|)} / \hbar}$. 

\section{Bremsstrahlung emission in alpha decay}
Given the number of theoretical works which have appeared on this subject 
over the years (as listed in the introduction too) the question that 
probably comes to the reader's mind here is: why are we proposing 
yet another model? We therefore begin by stating the reasons for such an 
undertaking. To start with, (i) the quantum time concepts were successfully 
applied to realistic examples in nuclear and particle physics such as 
locating particle resonances 
\cite{wetimedelay}, eta-mesic nuclear states \cite{meprl}, half lives of 
heavy nuclei and even in other branches like atomic, semiconductor physics, 
chemistry and biology (see \cite{us1} and references therein). 
It is certainly interesting to extend these concepts to an intriguing 
phenomenon in nuclear physics. (ii) The quantum mechanical treatments are 
based on the evaluation of the transition matrix involving integrals where 
a separation of the space regions before, within and after the barrier 
where the photon could have been emitted is not so obvious. Besides, while 
some papers simply use a rectangular well nuclear potential 
\cite{papen,takigawa}, others exclude 
the inner (nuclear potential) region from the integration \cite{giard1,madan}. 
The present work will use a realistic nuclear potential (with a double 
folding model of nuclear densities and the M3Y nucleon-nucleon interaction 
\cite{satchler, us2}) and verify the role of emission in the various spatial 
regions. (iii) Another new input is that the 
alpha-daughter cluster preformation probability is 
incorporated in the calculation and found to be important. 

\subsection{The semi-classical model} 
We begin by defining an average 
velocity of the particle between points $b$ and $a$ as 
\begin{equation}
< v > = {\int_{a}^{b} |\Psi(x)|^2 \, v(x)\, dx \over 
\int_{a}^{b} |\Psi|^2 \, dx} \, .
\end{equation}
With the wave function being stationary and hence the density 
$\rho = |\Psi|^2$ being time independent, the continuity equation is 
$\vec{\nabla} \cdot \vec{j} = 0$ and the current density $j$ 
is constant in the one dimensional problem. 
Identifying $j = \rho v$ in the above equation, 
\begin{equation}\label{avevel}
< v > = {j\, (b - a) \over 
\int_{a}^{b} |\Psi|^2 \, dx} \, 
= \, {b - a\over \tau_D}\, .
\end{equation}
Given the fact that we are interested in only those events where 
the alpha particle was transmitted through the barrier, we choose 
the constant flux $j$ to be the transmitted flux 
$j_T = \hbar \, k_0 |T|^2/ \mu$. In a semi-classical picture one 
could consider $b - a$ as the distance travelled by the particle 
while it spent the time $\tau_D$ in that region. 
Coming back to the alpha-nucleus potential one could then 
write this distance as the one between the classical turning points 
times the number of assaults, ${\cal N}$, made by the particle before leaving 
that region. 
For example, for the potential with the classical turning points 
$r_1$, $r_2$ and $r_3$ defined by $V(r) = E$ (where $E$ is the energy of 
the tunneling particle), 
the frequency of assaults at the
barrier, $\nu$, can be written as the inverse of the time required
to traverse the distance back and forth between the turning points $r_1$ and
$r_2$ as \cite{froeman},
\begin{equation}\label{period}
\nu\,=\,{\hbar \over 2\,\mu}\,\biggl[\,\int_{r_1}^{r_2}\,
{dr \over k(r)}\,\biggr]^{-1}\,.
\end{equation}
which is the inverse of twice the traversal
time (\ref{travtime}) from $r_1$ to $r_2$. The number of assaults 
made by the alpha in region I is then, ${\cal N}_I = \nu_I \, \tau_D$. 
With $\nu_I = 1/ (2 \,\tau_{trav}^I)$,  
\begin{equation}
{\cal N_I} \,= \,{\tau_D^I \over 2 \, \tau_{trav}^I} 
\end{equation}
Replacing for $b - a$ with ${\cal N_I} (r_2 - r_1)$ in (\ref{avevel})
for region I and similarly with ${\cal N_{II}} (r_3 - r_2)$ for region II,    
the average velocity in regions I and II can be finally written as 
\begin{equation}
v_{I}\, =\, {r_2 - r_1 \over 2 \,\tau_{trav}^I}\, ,\, \, 
v_{II}\, =\, {r_3 - r_2 \over 2 \,\tau_{trav}^{II}}
\end{equation}
The velocity in region III, $v_{III}$ is simply the free velocity 
and is given by $\sqrt{2\,E_{\alpha}/ \mu}$. 
Defining the times at the turning points $r_2$ and $r_3$ as 
$t_2$ and $t_3$ respectively, the velocity function can be written as 
\begin{equation}\label{fullvel}
v(t) \, =\, v_{I}\, \Theta(t_2 - t)\, + \, v_{II} \, \Theta(t_3 - t) 
\, \Theta(t - t_2) \, + \, v_{III} \, \Theta(t - t_3) 
\end{equation}
where the step function $\Theta(t_0 - t)$ is unity for all $t < t_0$ and 
zero otherwise. 
%Thus, 
%\begin{equation}
%{dv \over dt} = - v_{I}\, \delta(t_2 - t) + v_{II} \,
%[\delta(t-t_2) \Theta(t_3 -t) - \delta(t_3 -t) \Theta(t - t_2)]\, + \, 
%v_{III} \delta(t - t-3)
%\end{equation}
%and 

The classical formula for the photon emission probability in alpha decay 
is given as \cite{papen,dyak1}, 
\begin{equation}\label{prob}
{dP\over dE_{\gamma}} \, =\, P_{\alpha}\, 
{2 \alpha Z_{eff}^2 \over 3 \pi \,E_{\gamma}} 
\, |a_{\omega}|^2
\end{equation}
where 
\begin{equation}\label{aomega}
a_{\omega}\, =\, \int_{-\infty}^{\infty}\, dt \, {dv\over dt} \, 
e^{-i \omega t}
\end{equation}
and we have introduced a factor $P_{\alpha}$ in order to account for the 
alpha cluster preformation probability. 
$Z_{eff}$ is the effective charge for dipole transitions and is given 
as $Z_{eff} = (2A - 4Z)/(A + 4)$ where $A$ and $Z$ are the mass and 
atomic numbers of the daughter nucleus. For example, $Z_{eff}$ = 0.4 for 
$^{210}$Po decay. 
Replacing for the velocity from (\ref{fullvel}) in (\ref{aomega}) 
we obtain, 
\begin{equation}
a_{\omega}\, =\, [v_{II}(Q - \hbar \omega) - v_{I}(Q)] \, e^{-i\omega t_2}\,+
\, [v_{III}(Q - \hbar \omega) - v_{II} (Q)]\, e^{-i\omega t_3}
\end{equation}
where we have written the energy dependence of the velocities explicitly. 
$Q$ is the $Q$-value of the decay and $\hbar \omega$ is the energy of the 
emitted photon. 
This dependence appears due to the fact that energy conservation has to be 
respected (neglecting however the tiny recoil of the nucleus). The energy 
in $v_{III}$ should actually be $E_{\alpha} - \hbar \omega$, however, for 
all practical purposes, this does not lead to a big difference in the results. 
%\begin{eqnarray}\label{accel}
%|a_{\omega}|^2 \, =\, (v_{II}(Q - \hbar \omega) - v_{I}(Q))^2 + 
%(v_{III}(Q - \hbar \omega) - v_{II} (Q))^2 + \\ \nonumber
%2 \, (v_{III} (Q - \hbar \omega) - v_{II}(Q)) 
%(v_{II}(Q - \hbar \omega) - v_{I}(Q)) \, \cos{[\omega (t_3 - t_2)]} 
%\end{eqnarray}
$t_3$ and $t_2$ define the times at which the particle enters and 
leaves the barrier. We choose $t_3 - t_2$ in the interference term 
to be the traversal time in the 
barrier. Thus for a given alpha-nucleus potential, the velocities and 
hence $a_{\omega}$ can be calculated from the traversal times.
Evaluating the dwell times (and hence half life) \cite{us1}, the preformation 
factor is fixed (see the discussion below) 
and finally the emission probability is determined from 
(\ref{prob}).

\subsection{Potential and cluster preformation factor}
Starting with the standard definition of the WKB decay width \cite{gurvitz},
\begin{equation}\label{gurwidth}
\Gamma (E)\,=\,P_{\alpha} \, \,
{\hbar^2 \over 2 \,\mu}\,\,\biggl[\,\int_{r_1}^{r_2}\,
{dr \over k(r)}\,\biggr]^{-1}\,e^{-2\int_{r_2}^{r_3}\, \kappa(r)\,dr} \, , 
\end{equation}
where,  $k(r)\,=\,{\sqrt{ 2\mu\,(E\,-\,V(r))} / \hbar}$ and
$\kappa(r)\,=\,{\sqrt{ 2\mu\,(V(r)\,-\,E)} / \hbar}$, the half life of the 
nucleus can be evaluated to be $\tau_{1/2} \, =\, 
\hbar \, {\rm ln \,2} / \Gamma$.  
The factor $P_{\alpha}$ is determined by comparing the 
experimental half life of the nucleus with the theoretical one. 
The potential, 
$V(r)\, =\, V_n(r)\,+\,V_c(r)\,+\,{\hbar^2\,(l\,+\,1/2)^2 \over \mu\,r^2}$, 
where $V_n(r)$ and $V_c(r)$ are the nuclear and Coulomb parts of the 
$\alpha$-nucleus (daughter) potential, $r$ the distance between
the centres of mass of the daughter nucleus and alpha and $\mu$
their reduced mass. The last term
represents the Langer modified centrifugal barrier \cite{langer}. With the
WKB being valid for one-dimensional problems, the above modification from
$l(l+1) \,\rightarrow \,(l+1/2)^2$ is essential to ensure the correct
behaviour of the WKB scattered radial wave function near the origin as well as
the validity of the connection formulas used \cite{morehead}.
Another
requisite for the correct use of the WKB method is the Bohr-Sommerfeld
quantization condition, which for an alpha with energy $E$ is given as,
\begin{equation}\label{bohrsomer}
\int_{r_1}^{r_2} \,\, K(r)\,dr\,=\,(n\,+\,1/2)\,\pi
\end{equation}
where $K(r) \,=\, \sqrt{{2\mu \over \hbar^2}\,|V(r)\,-\,E|}$ and $n$ is the
number of nodes of the quasibound wave function of $\alpha$-nucleus
relative motion. 
The number of nodes are re-expressed as $n\, = \,(G\,-\,l)\,/2$, where
$G$ is a global quantum number obtained from fits to data \cite{chin,buck}. 
We choose $G = 22$ for the $^{210}$Po calculations.
The folded nuclear potential is written as,
\begin{equation}\label{potnucl}
V_n(r)\,=\,\lambda \,\int\,d{\bf r}_1\,d{\bf r}_2\,
\rho_{\alpha}({\bf r}_1) \, \rho_d({\bf r}_2)\,v({\bf r}_{12}\,=\,{\bf r}\,
+\,{\bf r}_2\,-\,{\bf r}_1,\,E)
\end{equation}
where $\rho_{\alpha}$ and $\rho_{d}$ are the densities of the alpha and the
daughter nucleus in a decay and $v({\bf r}_{12},E)$ is the nucleon-nucleon
interaction. $|\bf{r}_{12}|$ is the distance between a nucleon in the alpha
and a nucleon in the daughter nucleus. $v(\bf{r}_{12},E)$ is written using the
M3Y nucleon-nucleon (NN) interaction as in \cite{satchler}. 
The Coulomb potential is obtained using a similar double folding procedure
\cite{us2} 
with the matter densities of the alpha and the daughter replaced by their
respective charge density distributions $\rho^c_{\alpha}$
and $\rho^c_{d}$.

\subsection{Photon emission probabilities}
The photon emission probabilities evaluated within the semi-classical 
tunneling time model are presented in Figure 1 for the alpha decay of 
the nucleus $^{210}$Po. One can see that the contribution to the results 
from the acceleration at the beginning of the Coulomb barrier (dashed line) 
is much larger 
than the acceleration while leaving the barrier (dot-dashed line) . 
The shape of the total 
emission probability (solid line) however gets decided by the sum and 
interference of the two terms.
The disagreement with data (which as such also disagree with each other 
having three different slopes) 
at high energies 
could either be a limitation 
of the semi-classical model or due to the energy dependence of the 
cluster preformation factor (which in the present work has been 
chosen to be constant). 
It is also important to note that we obtain 
$P_{\alpha}$=0.03 on comparing the experimental and theoretical 
half lives of $^{210}$Po and this factor is essential to reproduce the right 
order of magnitude of the photon emission probability.
\begin{figure}[h]
\includegraphics[width=9cm,height=9cm]{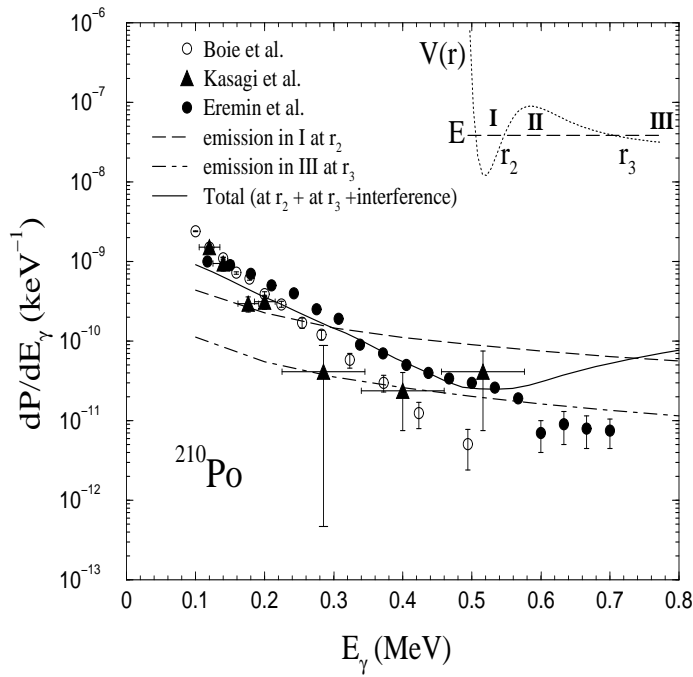}
\caption{\label{fig:eps2} Emission probabilities for bremsstrahlung 
accompanying the $\alpha$ decay of $^{210}$Po. The data are from 
Refs. \cite{kasagi1,boie,eremin}.}
\end{figure}

In order to test the sensitivity of the results to the potential used, in 
Figure 2 we display the results evaluated using the 
realistic potential $V(r)$ mentioned in the previous section and a 
simpler potential of the form, 
$V(r) = [2Z \alpha /r] \Theta(r - r_0) \, -\, V_0 \Theta(r_0 - r)$
where $V_0$ and $r_0$ are chosen to take the values used in \cite{papen} for 
$^{210}$Po. Using $V_0 =16.7$ MeV and $r_0 = 8.76$ fm as in \cite{papen}
and the $Q$ value of 5.407 MeV, the 
experimental half life in (\ref{gurwidth})
can be reproduced only after the inclusion of 
$P_{\alpha}$ = 0.03. One can also rewrite the rectangular potential as, 
$V(r) = [2Z \alpha /r] \Theta(r - r_0) \, -\,\lambda\, 
\tilde{V}_0 \Theta(r_0 - r)$ 
and adjust $\lambda$ in order to satisfy the Bohr-Sommerfeld condition. 
This leads to $V_0 = \lambda \tilde{V}_0$ = 75 MeV. 
It is interesting to see that such a 
rectangular well brings the results closer to those with the realistic 
potentials. The preformation factor however changes to $P_{\alpha}$ =0.016. 
\begin{figure}[h]
\includegraphics[width=12cm,height=9cm]{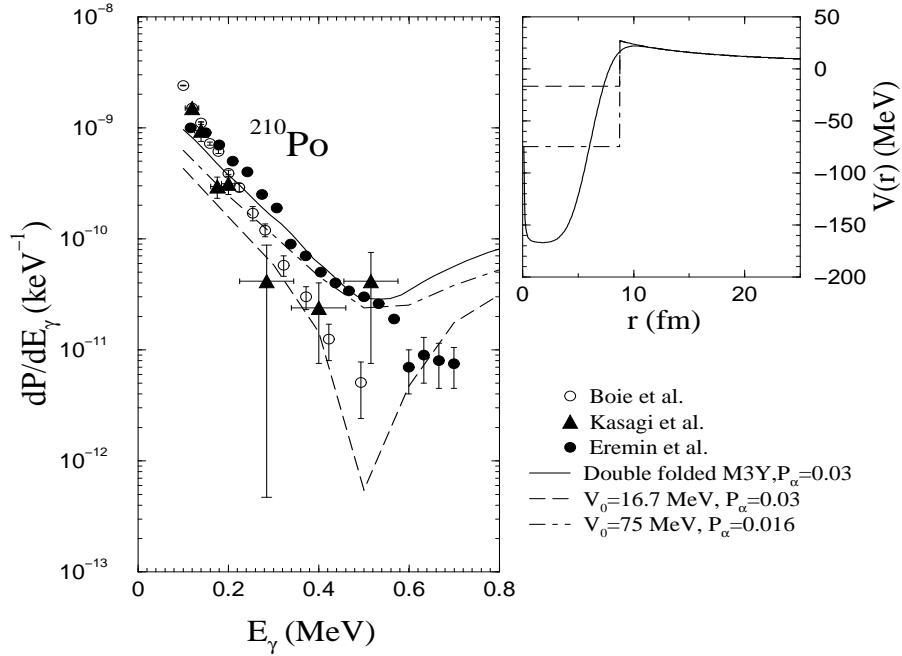}
\caption{\label{fig:eps3} Sensitivity of the photon emission probability
to the nuclear potential.}
\end{figure}

The semi-classical tunneling time model could in principle be applied to 
other existing data on the decay of $^{214}$Po, $^{226}$Ra and 
$^{244}$Cu. These results are not presented here since the qualitative 
behaviour of the emission probabilities remains the same. The magnitude of the 
results is sensitive to the input of the preformation factor which in turn 
gets decided by the strength of the nuclear potential (which is decided 
by the global quantum number input). For an input $G =24$ 
for example, the probabilities 
for $^{226}$Ra and $^{214}$Po are slightly overestimated 
as compared to data in the present approach. 

\section{Critical view of the theoretical approaches}

Apart from the fact that the data on bremsstrahlung emission in alpha 
decay are sparse, there exist contradictory conclusions from theoretical 
approaches in literature. In the present section we try to give an 
overview of the results from different approaches and a comparison of their 
conclusions. 
\subsection{Semi-classical approaches}
One of the first papers which appeared on this topic was that by Dyakonov 
and Gornyi \cite{dyak1} where the authors considered the tunneling motion of 
a charged particle using the semi-classical WKB wave functions. They derived 
a classical formula for the radiation spectral density in terms of the 
quantum mechanical traversal time delay $\Delta t$ which was given by, 
\begin{equation}
{\partial E \over \partial \omega} = {2 \over 3 \pi} {e^2 \over c^3} \, 
\omega^2 \, v_0^2 \, |\Delta t|^2 \, ,
\end{equation}
where the traversal time delay $\Delta t = \Delta t (-\infty)$ was
defined as the difference of the traversal time under the barrier and the 
free traversal time in the same region. 
The above spectral density is related to the experimentally measured 
emission probability by a factor proportional to $(4 \pi E_{\gamma})^{-1}$
\cite{dyak2}. 
The acceleration obtained in 
\cite{dyak1}, $|a_{\omega}^{DG}|^2 = \omega^ 2 v_0^2 |\Delta t|^2$ 
can be rewritten 
in terms of the average velocities appearing in the present work. 
Considering the fact that the authors in \cite{dyak1} consider a free $\alpha$ 
particle tunneling the barrier, the only contribution to the ``delay" is 
finite for the region within the barrier and elsewhere $\Delta t$ = 0. Thus, 
$\Delta t$ of Eq.(10) in \cite{dyak1} 
can be rewritten as,
\begin{equation}
\Delta t = \int_{r_2}^{r_3} \, {1 \over v(z)} dz \, - \, 
{r_3 - r_2 \over v_{III}}\, , 
\end{equation}
leading to $|a_{\omega}^{DG}|^2 = \omega^2 (\tau_{trav}^{II})^2 \, 
(v_{III} - 2v_{II})^2 $. This appears somewhat 
similar to our expression 
where if we retain the contribution only from the acceleration at the end 
of the barrier, we would obtain
$ |a_{\omega}|^2 \, =\,(v_{III} - v_{II})^2$. 
One would however expect $|a_{\omega}^{DG}|^2$ to grow with increasing photon 
energy as compared to $|a_{\omega}|^2$ of the present work. 
Working within the approach of \cite{dyak1} but with a different 
formalism \cite{dyak2} to evaluate $|a_{\omega}|^2$, Dyakonov obtained 
exponentially falling emission probabilities in reasonably good agreement 
with the $^{210}$Po data. 

The discrepancy to be noted here is that 
(i) Kasagi {\it et al}. \cite{kasagi1} obtained an almost perfect agreement 
with data (with a dip around $E_{\gamma}$ = 300 MeV), using the model 
proposed in \cite{dyak1}, (ii) the arguments presented above 
for $|a_{\omega}^{DG}|^2$ seem to suggest 
that it would be difficult to expect steeply falling probabilities with the 
expression in \cite{dyak1} and (iii) the author of \cite{dyak1} 
using an apparently similar formalism did obtain 
exponentially falling probabilities 
in \cite{dyak2}, however, with the absence of the dip and 
in disagreement with the result in (i) \cite{kasagi1}. The author mentioned 
a possible reason for the 
disagreement to be the use of different cut-offs of the Coulomb potential 
chosen in \cite{dyak2} and \cite{kasagi1}. 

\subsection{Quantum mechanical treatments} 
A fully quantum mechanical description \cite{papen} 
of the photon emission accompanying 
alpha decay followed the early experiments and the 
semi-classical theoretical approaches in 
\cite{dyak1,dyak2}. The authors expressed the emission probability 
in terms of a transition matrix involving the radial wave functions 
$\Phi_i$ and $\Phi_f$ 
of the initial and final $\alpha$ respectively and treating the 
photon field in the dipole approximation. The matrix element 
$<\Phi_f|\partial_r V |\Phi_i>$ was evaluated using the following potential: 
$V(r) = [2Z \alpha /r] \Theta(r - r_0) \, -\, V_0 \Theta(r_0 - r)$. 
The parameters $V_0$ and $r_0$ were fitted to obtain a half life consistent 
with an expression obtained from wave function matching. The authors found 
that the main contribution to photon emission stems from Coulomb acceleration 
and only a small contribution 
arises from the tunneling wave function under the barrier. This is in 
contrast to the findings of \cite{takigawa} where the authors (in a similar 
kind of quantum mechanical approach involving the calculation of 
the transition matrix elements with a rectangular nuclear potential) found 
the total spectrum to be a result of the interplay between different regions. 
The authors in \cite{takigawa} replaced the quantum mechanical Coulomb 
wave functions by semi-classical ones and divided the integral into 
different regions. They defined classical turning points and thus obtained 
semi-classical integral expressions for the tunneling, mixed and outside 
regions. Whereas Ref. \cite{papen} concluded that the soft-photon limit agrees 
with the classical results, Ref. \cite{takigawa} found classical theories 
inadequate in reproducing the subtle interference effects. In another 
quantum mechanical treatment \cite{taklya} 
of the interference of the different space regions in tunneling the results 
seemed to be in agreement with Ref. \cite{papen}.

A revived interest in the topic was seen by some more recent works 
\cite{giard1,giard2,madan} which studied the experimental spectra for 
photon emission accompanying the $^{210,214}$Po and $^{226}$Ra alpha decay. 
The authors in \cite{madan} for example employed a multipole expansion of 
the vector potential of the electromagnetic 
field of the daughter nucleus and also 
took into account the dependence on the angle between the directions of the 
$\alpha$ particle propagation and photon emission. They found the 
contribution of the photon emission during tunneling to be small. In 
their investigation of $^{226}$Ra they took into 
account the deformation of the 
nucleus and found the results to be different as compared to the spherically 
symmetric case. Even if they agreed in general with \cite{papen} that the 
tunneling motion contributes little, using the potential parameters of 
\cite{papen} they could however not reproduce the slope of the $^{210}$Po 
spectra.

\subsection{Time dependent formalisms} 
Finally, before ending this section we discuss two 
time dependent descriptions of the bremsstrahlung emission. 
In contrast to the stationary descriptions of quantum tunneling 
described so far, the authors in \cite{berty} resort to numerically 
solving the time-dependent Schr\"odinger equation. 
%The potential used 
%in solving the Schr\"odinger equation is somewhat similar to that of
%Refs. \cite{papen} and \cite{takigawa}, namely, a rectangular well for the 
%nuclear part and a $(1/r)$ Coulomb potential. 
The emission probability involves the radial momentum which is evaluated 
using the time dependent wave function. 
%Due to numerical difficulties posed 
%by the formalism the authors do not compare their results with data, however,
Apart from finding 
the time dependent modification of the wave function to be important, the 
authors notice that the usual assumption of a preformed alpha cluster 
in a well leads to 
sharp peaks at high frequencies in the bremsstrahlung emission. 
These peaks are interpreted as the manifestation of the fact that the 
initial localized state has some overlap from neighbouring resonant states.
Though the importance of these peaks would reduce if the initial state is a 
sharp resonance (as is the case for $^{210}$Po), the authors express the 
need for more experimental data on bremsstrahlung radiation by a tunneling 
particle in order to understand better the preformation of clusters and the 
above phenomenon of ``quantum beats".

In \cite{greiner} the authors propose a numerical algorithm based on the 
Crank-Nicholson method to solve the time dependent Schr\"odinger equation and 
thereby evaluate average position, momentum and acceleration in alpha decay. 
They conclude that a big effect of the tunneling motion should be expected in 
the region of hard photons. Though the authors do not compare their results 
with data, they find that the contribution coming from the tunneling motion is 
an order of magnitude smaller than that from Coulomb acceleration. 

\section{Summary}
To summarize the findings of the present work, we can say that:
\\
(i) We have presented a new semi-classical model based on the concept of 
quantum tunneling times in order to evaluate the photon emission 
probabilities in alpha decay of nuclei.
Special attention was paid to the use of realistic nuclear and Coulomb 
potentials and the results were found sensitive to the type of nuclear 
potential used.\\
(ii) A review of the existing theoretical literature shows that the opinion 
regarding the contribution of the photon emission during tunneling is divided 
among some who consider this motion as well as subtle interference effects 
between regions to be important and others who consider the Coulomb 
acceleration to be the dominant one.\\
(iii) The existing data on $^{210}$Po are not consistent 
with each other and for other nuclei
are few. 
We emphasize here the need for new reliable data in order to resolve 
the intriguing question which we started with: does the alpha particle 
emit radiation during tunneling?

\begin{acknowledgments}
The authors thank G. Giardina for his kind help with the bremsstrahlung 
data.
\end{acknowledgments}

\end{document}